\documentclass[sigconf,screen,nonacm]{acmart}
\AtBeginDocument{%
  \providecommand\BibTeX{{%
    \normalfont B\kern-0.5em{\scshape i\kern-0.25em b}\kern-0.8em\TeX}}}

\usepackage{subfigure}
\usepackage{multirow}
\usepackage{graphicx}

\begin{document}

\title[]{Recording and replaying psychomotor user actions in VR }

\author{Manos Kamarianakis}
\email{kamarianakis@uoc.gr}
\orcid{0000-0001-6577-0354}
\affiliation{%
  \institution{FORTH - ICS, University of Crete, ORamaVR}
  \country{Greece}
}

\author{Ilias Chrysovergis}
\orcid{0000-0002-5434-2175}
\email{ilias.chrysovergis@oramavr.com}
\affiliation{%
  \institution{ORamaVR}
  \country{Greece}
}

\author{Mike Kentros}
\orcid{0000-0002-3461-1657}
\email{mike.kentros@oramavr.com}
\affiliation{%
  \institution{FORTH - ICS, University of Crete, ORamaVR}
  \country{Greece}
}

\author{George Papagiannakis}
\orcid{0000-0002-2977-9850}
\email{papagian@ics.forth.gr}
\affiliation{%
  \institution{FORTH - ICS, University of Crete, ORamaVR}
  \country{Greece}
}

\renewcommand{\shortauthors}{Kamarianakis, Chrysovergis, Kentros, et al.}


\maketitle

\section{Introduction}

Nowadays, session recording and playback of a single or 
multi-user VR session has become an increasingly  market-required asset.
The need for effective VR recording and replaying (VRRR) is 
especially highlighted in virtual training applications, as replaying user actions can serve as an additional
and powerful educational tool.
Despite the effort, achieving VRRR is a task not natively 
undertaken by modern game engines and therefore most VR 
applications do not include such a feature by default. 

Current bibliography contains numerous studies of how the VR record and 
replay features can enhance the learning impact that VR educational-oriented applications  provide, 
by mainly measuring the performance of users \cite{Lahanas2015}.
Usually, the data are captured  in video format \cite{Zia2016} and 
a post process of this high-dimensional data is required to 
obtain any further analysis. 
Since video data is not sufficient for reasons we explain in Section~\ref{sec:our_approach}, our approach is close to Kloiber et al. \cite{Kloiber2020}, who proposed an analysis of user's motion by recording their hands and head trajectories. Current ongoing research also 
explores the proper methods 
and data structures that must be employed to achieve 
real-time logging while keeping the required data storage  manageable and allowing effective replay.

\begin{figure}
    \centering
    \includegraphics[height=50mm]{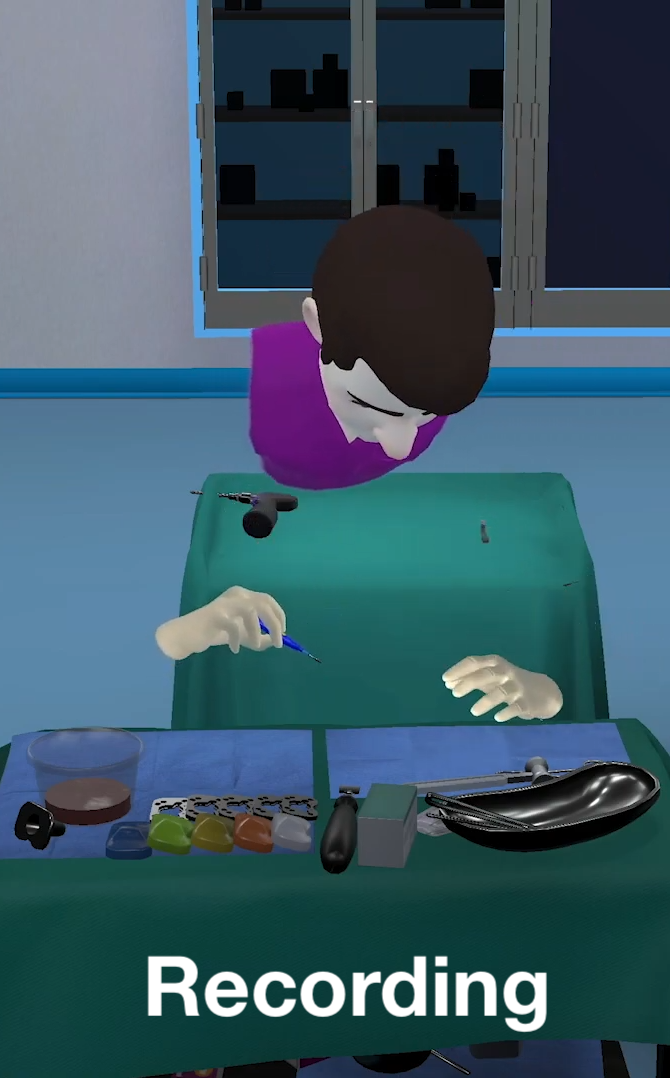}
    \includegraphics[height=50mm]{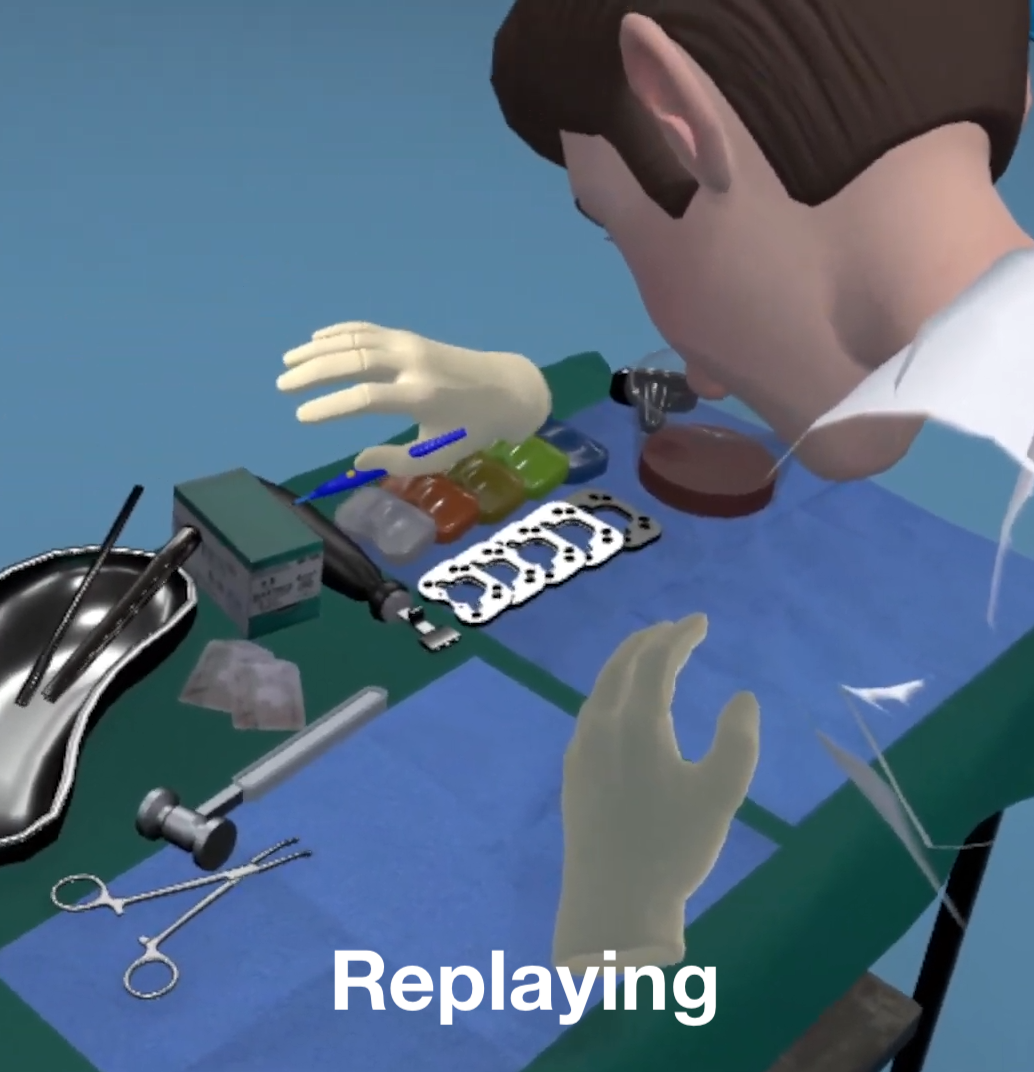}
    \caption{VR Record (Left) and Replay (Right) functionality.  The user is able to replay the recorded session from any perspective, as well as pause the replay in order to continue the session on his own.}
    \Description{VR Record and Replay functionality. }
    \label{fig:replaying}
\end{figure}

\section{Our Approach}
\label{sec:our_approach}
\textbf{\textit{Overview:}}
Our proposed VRRR functionality, already implemented in 
the Unity3D MAGES SDK 4.0 (\cite{papagiannakis2020mages}, publicly available for free), enables a) experts to record and replay 
their sessions, b) novices to learn how to correctly perform an operation 
by watching the expert's recording and reviewing their own sessions, 
and c) evaluators to assess the learning outcomes of the apprentices, utilizing the VR Replay feature (see Figure~\ref{fig:replaying}).

\textbf{\textit{VR Recorder:}} Existing methods dictate that accurate recording of a VR session can be achieved via two methods, depending on the logged data.
The first method records all users{\color{blue}'} inputs whereas, the second method
focuses on the effects that these inputs have on the virtual scene.
Our research revolves around a VR logger that records raw user input, as such an approach fits well with the
current growth direction of virtual reality environments. 
In this respect, we avoid collecting high-dimensional data, such as the state of the entire virtual world. Instead, we collect and store low-dimensional information, such as the users' inputs and triggers, that may be used to
obtain valuable analytics even by using simple, machine learning (ML) based, processing algorithms. Finally, the recorded data can be utilized to easily reproduce the user session by applying the world's mechanics, while 
the second method would demand sophisticated computer vision algorithms.

Our VR Recorder allows, for the first time, to log the user's sessions within a virtual environment in the form of positions, rotations, user-object interactions and training scenegraph nodes \cite{zikasVirtualRealityMedical2022a}, resulting in improved accuracy without compromising generalization,
while requiring minimal storage space, approximately 1MB per minute per user.
As the objects which the user interacts with might come into 
contact or interaction with other objects, i.e., changing their
location or status, we are also recording the transformations of all subsequently affected objects. 
The voice of each player is recorded individually, including 
incoming voice in cases of multi-player sessions.

\textbf{\textit{VR Replay:}} Via VRRR, we can replay a VR session, 
in both single and multi player modes. 
These recordings can be synchronized with the cloud and also be replayed on any device regardless of the original hardware they were recorded on.
This functionality is not just a video recording of the 
in-game view, but rather a full reproduction of the session 
as it happened when it was recorded. 
While replaying the session, the users are free to move around the virtual world, watch the scene from any angle, or act simultaneously with the 
various recorded interactions and events. Such 
a functionality allows the creation
of high fidelity VR replays that can be used creatively 
to increase the pedagogical benefits of various simulations.

\textbf{\textit{Audio-Video Synchronization:}}
Audio Video synchronization (AV-sync) is a common issue when 
one tries to replay graphics with audio. 
According to the European Broadcasting Union, 
the relative timing between audio and vision components 
must be less than 40 ms for sound before graphics, and 
less than 60ms for sound after graphics. 
The proposed AV-sync method is similar to the one presented by Zhang et. al \cite{Zhang2017}, which uses timestamps to eliminate the lip-sync error.

In this regard, two issues need to be tackled a) the graphics 
replay times and b) the remote user's 
sound reception for multiplayer sessions. 
For both cases, the local     
user's recording time is used as the baseline. 

The first issue is caused by the different 
frame rate between recording and replaying 
a session. As a result, a desynchronization between graphics and the respective sounds is 
created. The \emph{sound after graphics} issue is resolved by waiting the amount of frames needed to visualize graphical changes. On the other hand, the \emph{sound before graphics} is tackled by skipping visual changes.

The second issue is caused by the reception time difference 
between transformation and sound data, of a remote user. 
Usually, due to network anomalies, the local user receives the remote users' sound data in a non-constant rate
compared to the transformation data, which, in return, 
creates lack or excess of sound samples per frame.
This issue is eliminated by adding noise or removing samples respectively, to match the expected sound sample rate, i.e., $48000$ samples per second.

\balance

\begin{figure}[htb]
    \centering
    \includegraphics[width=0.45\textwidth]{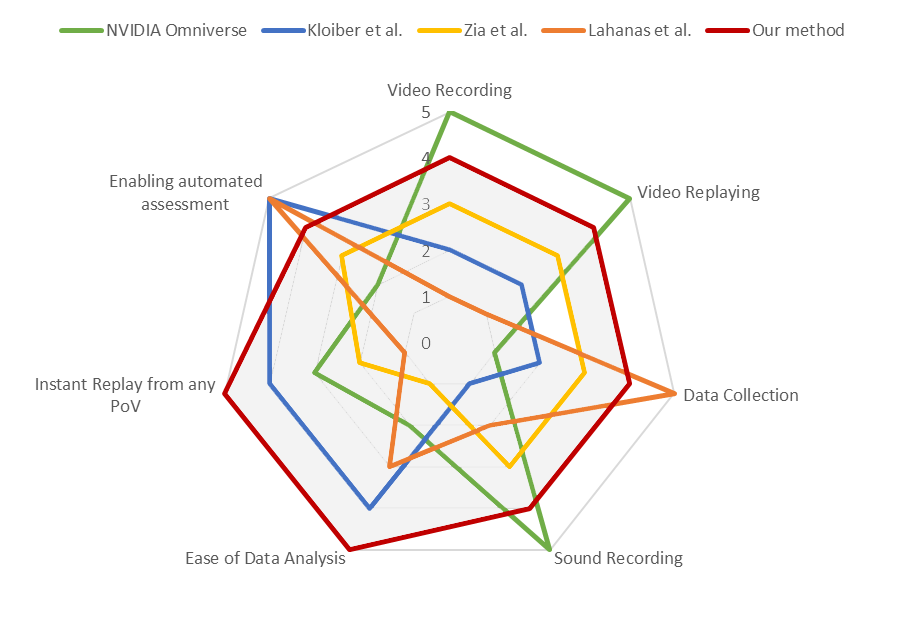}
    \caption{Comparison chart of existing VR recording and/or
    replaying methods \cite{Zia2016,Lahanas2015,Kloiber2020} against our proposed method.}
    \Description{Comparison chart of existing VR recording or/and
    replaying methods}
    \label{fig:comparison}
\end{figure}

\textbf{\textit{Results:}} A comparison chart of existing VR recording or/and 
replaying methods against our proposed method is presented in 
Figure~\ref{fig:comparison}. 
Moreover, Table~\ref{tbl:recorder_performance} shows a negligible performance overhead in terms of FPS in the use of our proposed VR Recorder.

\begin{table}[htb]
  \caption{Measuring the FPS burden of a VR application due to 
  the Recorder feature. The results were obtained using a PC with an i7-11375H, 16 GB of RAM and an RTX 3060.}
  \begin{center}
  \begin{tabular}{|c|c|c|c|}
  \hline
  \multirow{2}{*}{Metric}  & Session without & Session with & \multirow{2}{*}{Difference} \\
  & VR Recording & VR Recording &  \\
  \hline
  \hline
  Average FPS & 89.56  & 85.13  & 4.43 \\
  \hline
  Minimum FPS& 76.56  & 68.78  & 7.78 \\
  \hline
  Maximum FPS& 93.29  & 92.57  & 0.72 \\
  \hline 
  \end{tabular}
  \label{tbl:recorder_performance}
  \end{center}
\end{table}

\section{Conclusions and Related Work} 
\label{sec:conclusions}

We introduce a novel method that describes the functionality and characteristics of an efficient VR recorder with replay capabilities, implemented in a modern game engine, publicly available for free. 

Relating to this work, we have also developed a  
Convolution Neural Network (CNN) that can digest the stored data 
to assess in real-time the user's achievements
compared to prerecorded actions. In the future, 
we aim to create intelligent agents  trained using experts' session data, that are able to complete tasks on their own. Furthermore, we plan to develop a no-code VR
authoring tool, that would allow the virtual environment designers to develop new training modules.
This tool will also fuse 
all recorded data and train ML algorithms in order 
to understand the type of task the designer intends to develop.


\begin{acks}\label{sec:acks}
The project was partially funded by the European Union’s 
Horizon 2020 research and innovation programme under grant agreements 
No 871793 (ACCORDION) and No 101016509 (CHARITY).
We would like to thank Antonis Protopsaltis for his valuable comments.
\end{acks}

\bibliographystyle{ACM-Reference-Format}
\bibliography{references}


\begin{thebibliography}{6}


\ifx \showCODEN    \undefined \def \showCODEN     #1{\unskip}     \fi
\ifx \showDOI      \undefined \def \showDOI       #1{#1}\fi
\ifx \showISBNx    \undefined \def \showISBNx     #1{\unskip}     \fi
\ifx \showISBNxiii \undefined \def \showISBNxiii  #1{\unskip}     \fi
\ifx \showISSN     \undefined \def \showISSN      #1{\unskip}     \fi
\ifx \showLCCN     \undefined \def \showLCCN      #1{\unskip}     \fi
\ifx \shownote     \undefined \def \shownote      #1{#1}          \fi
\ifx \showarticletitle \undefined \def \showarticletitle #1{#1}   \fi
\ifx \showURL      \undefined \def \showURL       {\relax}        \fi
\providecommand\bibfield[2]{#2}
\providecommand\bibinfo[2]{#2}
\providecommand\natexlab[1]{#1}
\providecommand\showeprint[2][]{arXiv:#2}

\bibitem[Kloiber et~al\mbox{.}(2020)]%
        {Kloiber2020}
\bibfield{author}{\bibinfo{person}{Simon Kloiber}, \bibinfo{person}{Volker
  Settgast}, \bibinfo{person}{Christoph Schinko}, \bibinfo{person}{Martin
  Weinzerl}, \bibinfo{person}{Johannes Fritz}, \bibinfo{person}{Tobias
  Schreck}, {and} \bibinfo{person}{Reinhold Preiner}.}
  \bibinfo{year}{2020}\natexlab{}.
\newblock \showarticletitle{Immersive analysis of user motion in {VR}
  applications}.
\newblock \bibinfo{journal}{\emph{The Visual Computer}} \bibinfo{volume}{36},
  \bibinfo{number}{10-12} (\bibinfo{date}{Aug.} \bibinfo{year}{2020}),
  \bibinfo{pages}{1937--1949}.
\newblock


\bibitem[Lahanas et~al\mbox{.}(2015)]%
        {Lahanas2015}
\bibfield{author}{\bibinfo{person}{Vasileios Lahanas},
  \bibinfo{person}{Constantinos Loukas}, \bibinfo{person}{Nikolaos Smailis},
  {and} \bibinfo{person}{Evangelos Georgiou}.} \bibinfo{year}{2015}\natexlab{}.
\newblock \showarticletitle{A novel augmented reality simulator for skills
  assessment in minimal invasive surgery}.
\newblock \bibinfo{journal}{\emph{Surg. Endosc.}} \bibinfo{volume}{29},
  \bibinfo{number}{8} (\bibinfo{year}{2015}), \bibinfo{pages}{2224--2234}.
\newblock


\bibitem[Papagiannakis et~al\mbox{.}(2020)]%
        {papagiannakis2020mages}
\bibfield{author}{\bibinfo{person}{George Papagiannakis}, \bibinfo{person}{Paul
  Zikas}, \bibinfo{person}{Nick Lydatakis}, \bibinfo{person}{Steve Kateros},
  \bibinfo{person}{Mike Kentros}, \bibinfo{person}{Efstratios Geronikolakis},
  \bibinfo{person}{Manos Kamarianakis}, \bibinfo{person}{Ioanna Kartsonaki},
  {and} \bibinfo{person}{Giannis Evangelou}.} \bibinfo{year}{2020}\natexlab{}.
\newblock \showarticletitle{MAGES 3.0: Tying the Knot of Medical VR}. In
  \bibinfo{booktitle}{\emph{ACM SIGGRAPH 2020 Immersive Pavilion}}.
  \bibinfo{publisher}{Association for Computing Machinery}, Article
  \bibinfo{articleno}{6}, \bibinfo{numpages}{2}~pages.
\newblock


\bibitem[Zhang et~al\mbox{.}(1705)]%
        {Zhang2017}
\bibfield{author}{\bibinfo{person}{Min Zhang}, \bibinfo{person}{Yu-Hua Wu},
  \bibinfo{person}{Ji Li}, {and} \bibinfo{person}{Shi-Jun Li}.}
  \bibinfo{year}{2017/05}\natexlab{}.
\newblock \showarticletitle{Research on Audio and Video Synchronization
  Algorithm Based on AVI Format}. In \bibinfo{booktitle}{\emph{Proceedings of
  the 2017 2nd International Conference on Materials Science, Machinery and
  Energy Engineering (MSMEE 2017)}}. \bibinfo{publisher}{Atlantis Press},
  \bibinfo{pages}{959--962}.
\newblock
\showISBNx{978-94-6252-346-3}
\showISSN{2352-5401}


\bibitem[Zia et~al\mbox{.}(2016)]%
        {Zia2016}
\bibfield{author}{\bibinfo{person}{Aneeq Zia}, \bibinfo{person}{Yachna Sharma},
  \bibinfo{person}{Vinay Bettadapura}, \bibinfo{person}{Eric~L Sarin},
  \bibinfo{person}{Thomas Ploetz}, \bibinfo{person}{Mark~A Clements}, {and}
  \bibinfo{person}{Irfan Essa}.} \bibinfo{year}{2016}\natexlab{}.
\newblock \showarticletitle{Automated video-based assessment of surgical skills
  for training and evaluation in medical schools}.
\newblock \bibinfo{journal}{\emph{Int. J. Comput. Assist. Radiol. Surg.}}
  \bibinfo{volume}{11}, \bibinfo{number}{9} (\bibinfo{date}{Sept.}
  \bibinfo{year}{2016}), \bibinfo{pages}{1623--1636}.
\newblock


\bibitem[Zikas et~al\mbox{.}(2022)]%
        {zikasVirtualRealityMedical2022a}
\bibfield{author}{\bibinfo{person}{Paul Zikas}, \bibinfo{person}{Steve
  Kateros}, \bibinfo{person}{Nick Lydatakis}, \bibinfo{person}{Mike Kentros},
  \bibinfo{person}{Efstratios Geronikolakis}, \bibinfo{person}{Manos
  Kamarianakis}, \bibinfo{person}{Giannis Evangelou}, \bibinfo{person}{Ioanna
  Kartsonaki}, \bibinfo{person}{Achilles Apostolou}, \bibinfo{person}{Tanja
  Birrenbach}, \bibinfo{person}{Aristomenis~K. Exadaktylos},
  \bibinfo{person}{Thomas~C. Sauter}, {and} \bibinfo{person}{George
  Papapagiannakis}.} \bibinfo{year}{2022}\natexlab{}.
\newblock \showarticletitle{Virtual {{Reality Medical Training}} for {{COVID-19
  Swab Testing}} and {{Proper Handling}} of {{Personal Protective Equipment}}:
  {{Development}} and {{Usability}}}.
\newblock \bibinfo{journal}{\emph{Frontiers in Virtual Reality}}
  \bibinfo{volume}{2} (\bibinfo{year}{2022}).
\newblock
\showISSN{2673-4192}


\end{thebibliography}

\end{document}